\documentclass[nonacm]{acmart}

\usepackage{multirow}
\usepackage{array}
\usepackage{graphicx}
\usepackage{soul}
\usepackage{enumitem}

\AtBeginDocument{%
  }

\begin{document}


\title{Beyond the Legal Lens: A Sociotechnical Taxonomy of Lived Privacy Incidents and Harms}

\author{Kirsten Chapman}
\email{kac273@byu.edu}
\affiliation{%
  \institution{Brigham Young University}
  \city{Provo}
  \country{U.S.A}
}
 
\author{Garrett Smith}
\affiliation{%
  \institution{Brigham Young University}
  \city{Provo}
  \country{U.S.A}
}

\author{Kaitlyn Klabacka}
\affiliation{%
  \institution{Brigham Young University}
  \city{Provo}
  \country{U.S.A}
}

\author{Harrison Winslow}
\affiliation{%
  \institution{Brigham Young University}
  \city{Provo}
  \country{U.S.A}
}

\author{Louise Barkhuus}
\affiliation{%
  \institution{New York University}
  \city{New York City}
  \country{U.S.A}
}

\author{Cori Faklaris}
\affiliation{%
  \institution{University of North Carolina at Charlotte}
  \city{Charlotte}
  \country{U.S.A}
}

\author{Sauvik Das}
\affiliation{%
  \institution{Carnegie Mellon University}
  \city{Pittsburgh}
  \country{U.S.A}
}

\author{Pamela Wisniewski}
\affiliation{%
  \institution{International Computer Science Institute}
  \city{Berkeley}
  \country{U.S.A}
}

\author{Bart Piet Knijnenburg}
\affiliation{%
  \institution{Clemson University}
  \city{Clemson}
  \country{U.S.A}
}

\author{Heather Lipford}
\affiliation{%
  \institution{University of North Carolina at Charlotte}
  \city{Charlotte}
  \country{U.S.A}
}

\author{Xinru Page}
\email{xinru@cs.byu.edu}
\affiliation{%
  \institution{Brigham Young University}
  \city{Provo}
  \country{U.S.A}
}

\renewcommand{\shortauthors}{Chapman et al.}

\begin{abstract}

To understand how privacy incidents lead to harms, HCI researchers have historically leveraged legal frameworks. However, these frameworks expect acute, tangible harms and thus may not cover the full range of human experience relevant to modern-day digital privacy. To address this gap, our research builds upon these existing frameworks to develop a more comprehensive representation of people’s lived experiences with privacy harms. We analyzed 369 privacy incidents reported by individuals from the general public. We found a broader range of privacy incidents and harms than accounted for in existing legal frameworks. The majority of reported privacy harms were not based on tangible harm, but on fear and loss of psychological safety. We also characterize the actors, motives, and information associated with various incidents. This work contributes a new framework for understanding digital privacy harms that can be utilized both in research and practice. 
\end{abstract} 


\received{25 November 2025}

\maketitle

\section{Introduction}
Human-Computer Interaction (HCI) research has found many instances where privacy violations lead to harm \cite{freed2023understanding, wu2023slow, shelby2023sociotechnical}. Most of this research  captures highly contextual accounts of privacy violations for a specific population and technology. Scholars do not yet have a more systematic way to explain the wide variance in people's privacy experiences with slight changes in context, populations, or technologies. For example, surveillance apps designed for the safety of certain groups (e.g., child monitoring apps) have unintended and severe harms for other populations such as victims of intimate partner violence or migrants \cite{chatterjee2018spyware, guberek2018keeping}. Moreover, while both of these populations may experience psychological distress from the threat of surveillance \cite{guberek2018keeping, havron2019clinical}, how this distress manifests and strategies for resolution may look very different: for the former, law enforcement may be the first line of defense, but for the latter law enforcement may be the key threat actor.  

Can we build on such highly contextualized understandings of lived privacy experiences to provide a common vocabulary for notionally similar experiences of incidents and harm across technologies and context? This lack of a common vocabulary makes broad regulation and policy difficult, especially as emerging technologies introduce new vectors for privacy harm (e.g., LLMs can memorize and distribute private information about people in their pre-training data \cite{carlini2022quantifying}, and can make sensitive inferences about people such as where they are located \cite{mendes2024granular, staab2023beyond}). It also makes it difficult for researchers, systems designers, and technology end-users to anticipate privacy implications of new technologies or existing technologies in new contexts or with new actors. 

Attempts to provide a common vocabulary across lived privacy incidents and harms have often been adapted from legal conceptions of privacy incidents and harm which are constructed based on U.S. case and tort law. Two widely-used and foundational frameworks are Solove's Taxonomy of Privacy and Solove and Citron's Typology of Harms \cite{solove2005taxonomy, citron2022privacy}. Yet, the results of empirical research point to how privacy harms are experienced very differently than harms found in tort law (that traditionally tend to be acute and obvious) \cite{wu2023slow}. For example, targeted advertising consists of multiple small violations over time. This can lead to unease and perceived creepiness \cite{ur2012smart}. Unlike a more tangible harm such as theft, it is difficult to quantify the amount of harm derived from targeted advertising.

While prior scholars have despaired that the concept of privacy is too broad in scope to ever define simply \cite{solove2002conceptualizing}, we hypothesize that the harms experienced through privacy violations, even across a breadth of technologies and contexts, may be more tractable to categorize. As such, we ask the following research questions:

\begin{quote}
    \textbf{RQ1: } When people describe privacy incidents, what are common characteristics of these incidents shared across technologies and contexts of use? \\
    \textbf{RQ2: } For the incidents described, how do people report and experience the resulting privacy harms? \\
    \textbf{RQ3: } How can we adapt Solove and Citron's taxonomies to account for how people experience privacy incidents and harms? 
\end{quote}

To answer these research questions, we deployed an online survey (N=164) in January of 2025. This survey collected data on the online privacy incidents and harms that participants had reported experiencing. We then analyzed whether these reported experiences aligned with Citron and Solove's taxonomies on privacy. We found that while participants did experience privacy incidents described in Solove's Taxonomy of Privacy, they additionally experienced a wide range of privacy incidents which were unaccounted for in the taxonomy \cite{solove2005taxonomy}. Similarly, we found that while many of the harms which participants experienced aligned with Solove and Citron's Typology of Privacy Harms \cite{citron2022privacy}, many others deviated from harms traditionally represented by tort law. Instead, participants predominantly reported harms that accumulate over time and lead to fatigue and distress, rather than harms that are acute or easily quantifiable. Furthermore, while we found an expanded set of actors and information types associated with privacy incidents, the harms mostly align with Citron and Solove's typology. In summary, our research makes the following contributions to the literature: 
\begin{itemize} [noitemsep, topsep=0pt]
    \item \textbf{Understanding the range of privacy incidents and harms.} From analyzing participant reported privacy experiences, we are able to explore the types of privacy incidents and harms which may be prominent for technology users today.
    \item \textbf{Adapting Solove's Taxonomy of Privacy. } Participants reported privacy incidents with a wider range of actors and mechanisms than previously defined. We propose an adaptation of Solove's privacy incident taxonomy that can encompass the wider range of privacy incidents and actors relevant for HCI research.   
    \item \textbf{Adapting Citron and Solove's Typology of Privacy Harms.} Participants reported harms currently unaccounted for by Citron and Solove's framework \cite{citron2022privacy}, such as experiencing loss of psychological safety. We adapt this framework to account for these new forms of harm. 
    \item \textbf{Guidelines for Researchers, Technology Designers, and Policy Makers.} Providing an updated taxonomy of privacy incidents and harm will allow researchers to better understand,  technology designers to better anticipate and design against, and policymakers to create more effective legislation against privacy incidents and harms. 
\end{itemize} 
In addition to making these contributions, we believe there are strong implications for conducting privacy research. We conclude our paper with a discussion of these implications and recommendations for researchers. 

\section{Background}
In this section, we detail prior work relevant to HCI, sociotechnical perspectives on privacy, and the use of legal frameworks within HCI.

\subsection{Socio-Technical Perspectives on Modern-Day Privacy}
Privacy is normative and driven by the sociopolitical climate in a given context \cite{wisniewski2022privacy, knijnenburg2022modern, proferes2022development}. Human-centered computing research has examined privacy concerns and issues in a wide range of domains (including social media \cite{page2022social, xu2008examining, ellison2011negotiating}, privacy enhancing technologies \cite{seamons2022privacy}, tracking and personalization \cite{khan2024teaching, zhang2014privacy, berkovsky2015web, bujlow2017survey}, healthcare technologies \cite{serrano2016willingness, o2004health, genaro2015overview}, IoT devices \cite{medaglia2010overview, tabassum2020smart, vitak2018privacy}, and AI \cite{lee2024deepfakes}) and audiences (including cross-cultural settings \cite{trepte2017cross}, adolescents \cite{wisniewski2018privacy, kim2025privacy, kumar2017no, badillo2021conducting}, and vulnerable populations \cite{page2022perceiving, murthy2021individually, mcdonald2021s}). These domains and audiences all have unique distinctions in the types of privacy incidents and harms that users may incur. Additionally, as technologies evolve, we see differences in how specific populations in particular contexts are being harmed through privacy violations. For example, AI systems used in hiring processes that utilize prospective employees' data have been found to discriminate against individuals with disabilities \cite{buyl2022tackling}. 

Due to its  multi-faceted and highly contextual nature, some scholars have doubted whether privacy can be adequately understood by a single theory or framework \cite{knijnenburg2022modern}. Currently, there is a wide array of privacy frameworks being used in HCI research. Some of these frameworks help researchers understand privacy at a conceptual level. For instance, Altman's Boundary Regulation framework describes privacy as a process individuals undertake to regulate the amount of access others have to them \cite{altman1975environment}. 
Other frameworks delve more into how privacy violations occur. For example, Nissenbaum's Contextual Integrity framework describes privacy violations using four dimensions: the social context in a given situation, involved actors (sender and recipient of information), type of information shared, and transmission principles of how information is being relayed between actors \cite{nissenbaum2009privacy, wisniewski2022privacy}. Subsequent work has tried to apply and adapt these frameworks to specific technological contexts. Researchers have extended boundary regulation to account for the specific features of a given platform \cite{karr2011new, page2019communication, stutzman2012boundary, lampinen2011we}. For instance, Wisniewksi et al., described five categories of boundary mechanisms pertinent to social networking sites \cite{karr2011new}. HCI work has also examined creating frameworks to address the harms that specific technologies may cause. For instance, Shelby et al., examined 172 publications regarding algorithmically driven harms. This paper found that these technologies led to representational, allocative, quality of service, interpersonal, and social system harms \cite{shelby2023sociotechnical}.

As technology and the sociopolitical landscape change continuously, we must continually examine the privacy frameworks we use, lest we risk missing important new types of privacy incidents and harms that emerge from these changes. This is one of the main drivers of the current paper.

\subsection{Solove and Citron's Taxonomies of Privacy} 
In comparison to the more abstract theories and context-specific taxonomies described in the previous section, Solove and Citron's taxonomies of privacy are more broadly applicable. These taxonomies are designed to be technology-agnostic and apply across contexts. Solove's Taxonomy of Privacy \cite{solove2005taxonomy} was developed in the early 2000s and grounded in U.S. legal regulation and case law. It focuses on the transfer of information and the role of government in privacy incidents. This framework conceptualizes four categories of privacy violations: (1) information dissemination, (2) information processing, (3) information collection, and (4) invasion. Information dissemination incidents occur when personal data or information is shared with others. The \textit{threat} of information being shared is also considered an information dissemination violation. Information processing privacy incidents occur when data that has already been collected is handled in an unexpected way. Specifically, Solove describes the, ``use, storage, and manipulation'' of data. Information collection incidents occur when data is gathered in a violating way. Note that collection incidents do not require the actual public revelation of this data. These categories draw upon the flow of data and information from an individual. Invasion on the other hand, does not necessarily involve data, and instead focuses on privacy violations that occur directly to an individual. These categories are broken down into 16 distinct types of incidents (see appendix~\ref{appendixFrameworks}).

Solove and Citron later developed a Typology of Privacy Harms \cite{citron2022privacy} which was published in 2021 with the objective to illuminate why certain harms should be considered cognizable in a court of law. This framework once again examines privacy from a legal perspective with more emphasis on provable, or tangible, harms. The typology includes 7 comprehensive types of privacy harm: (1) Physical Harms, (2) Economic Harms, (3) Reputational Harms, (4) Psychological Harms, (5) Autonomy Harms, (6) Discrimination Harms, and (7) Relationship Harms. Both Psychological Harms and Autonomy Harms are further broken down into subcategories. See appendix ~\ref{appendixFrameworks} for the types and definitions of privacy harms. 

The next section will demonstrate that the two Solove taxonomies have seen some adaption in the field of HCI. However, while some recent work points to the more tangible harms described in Solove and Citron's taxonomy (e.g., economic harms), HCI literature  more frequently focuses on intangible harms such as psychological, social, and political harms \cite{mun2024particip, keum2023benefits, acemoglu2021harms}. As the incidents and harms defined in the two taxonomies are based on (and constrained by) U.S. case and tort law, it is important to understand what additional incidents and harms users may experience. Our work examines whether privacy incidents and harms reported by people align with these legal frameworks.

\subsection{The Use of Solove and Citron's Frameworks in HCI}
Work in the CHI and CSCW community has utilized Solove's Taxonomy of Privacy \cite{solove2005taxonomy} and Solove and Citron's Privacy Harms framework \cite{citron2022privacy} to study privacy in specific contexts and domains. Much of this work has been in regards to AI-related privacy incidents and harms. Lee et al., examined the effects of AI on privacy incidents defined in Solove's Taxonomy of Privacy\cite{lee2024deepfakes}. By analyzing privacy incidents documented in the repository of AI, Algorithm, and Automation Incidents and Controversies (AIAAIC) \cite{AIAAIC}, these researchers found that AI exacerbates certain privacy incidents (e.g., surveillance) and creates new risks such as new types of distortion incidents (i.e., the creation of fake, yet realistic, audio and images). AI-related privacy incidents have also been shown to lead to additional harms currently unaccounted for in Solove and Citron's Privacy Harms framework. For example, emotion-AI has been found to lead to emotional labor harms, such as feeling the need to display certain emotions, for job seekers and workers \cite{pyle2024us, roemmich2023emotion}.

Other research on workers' online privacy has found further omitted harms. Research utilizing Solove's Taxonomy of Privacy to understand the violations MTurk workers experience, found that deceptive practices (e.g., scams, malware) were frequent \cite{xia2017our}. Studies in the context of gig workers found that the way Solove and Citron described autonomy harms diverged from how their participants felt these adverse effects. That is, they found that while Citron and Solove relate autonomy harms to an individual's inability to make informed decisions regarding their data, their participants were instead experiencing harms in relation to their offline, physical bodies \cite{rivera2024safer}. Other work has cited additional use-case boundaries they found with these frameworks. One study found that only physical, economic, and reputation harms applied to organizations, while all harms were applicable to individuals \cite{dev2023models}. 

The fact that HCI researchers have been finding additional privacy incidents and harms not currently included in the two frameworks indicates that these frameworks may need to be adapted to truly account for the nuances in HCI. Our work seeks to empirically examine how individuals experience privacy incidents and harms in order to understand how these frameworks can be adapted.  

\section{Methods}
To better understand the types of privacy incidents individuals experience, we developed and deployed an online survey. The rationale for using the survey method is that it allowed for the collection of experiences from a broad range of individuals from the general public \cite{lazar2017research}.

\subsection{Survey Instrument}
The survey contained the following parts: 
\begin{enumerate}
    \item \textbf{Informed Consent.} Participants shown an informed consent page regarding the survey.
    \item \textbf{Closed-Ended Questions about Privacy Incidents.} Participants were shown definitions of each of the 16 types of online privacy incidents described in Solove's taxonomy and asked to answer yes/no regarding whether they had experienced each of them \cite{solove2005taxonomy}. We grounded these questions in this well-known taxonomy because it covers a broad variety of incident types while simultaneously not focusing on specific technologies. The exact questions asked can be found in appendix ~\ref{survey}.
    \item \textbf{Open-Ended Questions about Three Randomly Assigned Incidents.} Participants were then randomly assigned 3 privacy incident categories from the subset of incidents they had answered 'yes' to in the previous step. If participants hadn't answered 'yes' to experiencing three or more of the types of privacy incidents (16\% of participants), they were automatically assigned the one or two incidents they had said 'yes' to, and then asked to share a hypothetical instance of the incident for one or two additional types of incidents (randomly assigned) to ensure the workload remained consistent. For the purpose of this study, data related to such hypothetical incidents were excluded from the analysis. These participants were asked to share the following about each incident: 
    \begin{itemize}
        \item What was the incident?
        \item What caused the incident?
        \item Where the incident occurred (social media, mobile app, etc.)
        \item How did the incident cause discomfort or harm?
    \end{itemize}
    The open-ended responses to these questions were analyzed in order to answer our research questions.
\end{enumerate}

\subsection{Recruitment}
Participants were recruited through the crowd-sourcing platform Prolific from January 23-30 in 2025. We recruited through Prolific as it has been found to provide high-quality and meaningful responses \cite{douglas2023data, peer2022data}. We restricted recruitment to U.S. participants since privacy attitudes can vary across cultures. Quota sampling was utilized with a 50:50 ratio between males and females, and a representative sample of household income. 

In total, 164 participants were recruited. 5 quality checks (including three close-ended attention checks, 1 open-ended attention check, and one reverse-coded question) were utilized to ensure authentic responses. No participants failed more than 1 of these quality checks, so all data was retained. 

Filtering out hypothetical violations left us with 151 participants (see table ~\ref{tab:demographics}) sharing an average of 2.76 open-ended responses (min: 1, max: 3) about privacy violations they had incurred, for a total of 369 responses. The median completion time for the survey was 15 minutes and 27 seconds. Individuals were paid \$3.00 USD for their participation in the study (a rate of \$12/hour).

\begin{table}[htp!]
\small
\begin{tabular}{l|lrr}
\toprule
\textbf{Attribute} & \textbf{Value} & \textbf{\#} & \textbf{\%} \\
\midrule
\multirow{4}{*}{Gender} & Female                       & 66  & (43.7\%) \\  
                        & Male                         & 82  & (54.3\%) \\  
                        & Other                        & 1   & (0.7\%)  \\  
                        & Prefer not to say            & 2   & (1.3\%)  \\ \midrule
\multirow{7}{*}{Age}    & 18 - 24                      & 32  & (21.2\%) \\  
                        & 25 - 34                      & 44  & (29.1\%) \\  
                        & 35 - 44                      & 29  & (19.2\%) \\  
                        & 45 - 54                      & 25  & (16.6\%) \\  
                        & 55 - 64                      & 11  & (7.3\%)  \\  
                        & 65+                          & 8   & (5.3\%)  \\  
                        & Prefer not to say            & 2   & (1.3\%)  \\ \midrule
\multirow{7}{*}{\begin{tabular}[c]{@{}l@{}}Race\\ (Select\\ all that\\ apply)\end{tabular}} & American Indian or Alaska Native        & 1  & (0.7\%)  \\  
                        & Asian                        & 8   & (5.3\%)  \\  
                        & Black or African American    & 24  & (15.9\%) \\  
                        & Hispanic or Latino           & 17  & (11.3\%) \\  
                        & White                        & 107 & (70.9\%) \\  
                        & Other                        & 5   & (3.3\%)  \\  
                        & Prefer not to say            & 1   & (0.7\%)   \\ \midrule
\multirow{6}{*}{Education}                                                                  & Graduated Secondary Education           & 28 & (18.5\%) \\  
                        & Some College/University      & 43  & (28.5\%) \\  
                        & Graduated College/University & 47  & (31.3\%) \\  
                        & Started Post-Graduate Degree & 4   & (2.6\%)  \\  
                                                                                            & Post-Graduate Degree (MS, PhD, JD, etc) & 26 & (17.2\%) \\  
                        & Prefer not to say            & 3   & (2\%)    \\ \midrule
\multirow{7}{*}{\begin{tabular}[c]{@{}l@{}}Household\\ Income\end{tabular}}                 & Less than \$15,000                      & 18 & (11.9\%) \\  
                        & \$15,000 to \$34,999           & 23  & (15.2\%) \\  
                        & \$35,000 to \$49,999           & 27  & (17.9\%) \\  
                        & \$50,000 to \$74,999           & 24  & (15.9\%) \\  
                        & \$75,000 to \$99,999           & 27  & (17.9\%) \\  
                        & \$100,000 or more            & 31  & (20.5\%) \\  
                        & Prefer not to say            & 1   & (0.7\%)  \\ \bottomrule
\end{tabular}
\caption{Participant Demographics}
\label{tab:demographics}
\end{table}

\subsection{Qualitative Analysis}
Two researchers independently coded the open-ended responses for privacy incidents using a theoretically driven approach supplemented by open coding to allow for emerging themes. They used Solove's Taxonomy of Privacy to code privacy incidents \cite{solove2005taxonomy}. The researchers met and discussed any differences until reaching consensus---a commonly accepted best practice in qualitative analysis \cite{mcdonald2019reliability}. This process was repeated to identify the prevalence of different types of harms, using Solove and Citron's Typology of Privacy Harms \cite{citron2022privacy}. 
Responses were coded based on the content written, not based on the original question (e.g. a response to a question about reputational harm that actually described relationship harm was coded as the latter). Where applicable, privacy experiences were coded to multiple types of incidents and harms (e.g., a reported experience could include both information dissemination and information collection incidents).

Our open coding led us to observe that 3 additional dimensions were relevant to understanding respondent's privacy experiences: information types, actors, and underlying motives of the instigating actor. One researcher created a codebook based on all of the responses, and then a second researcher independently used the codebook to code all responses. The researchers met and reconciled any differences through discussion until reaching consensus.

\subsection{Positionality Statement}
As HCI privacy researchers, we have both a technical and research-based understanding of online privacy that allowed us to differentiate between subtly different types of privacy incidents and harms. These differences were not always apparent to our participants who would report on a privacy incident different from their prompt. Nonetheless, our participants are the experts of their own lived experiences and shared a valid recounting of these experiences. Given our background, we were able to translate those experiences into the taxonomies for privacy incidents and harms. 

Given our focus on human experience, it is likely that we classify a much broader range of these experiences as privacy harms than would scholars in other disciplines. Legal scholars may want to only focus on the more tangible and quantifiable harms. While our broader focus allows us to extend the frameworks to be more broadly applicable in the field of HCI, we acknowledge that other scholars may find the original taxonomies or variations of this updated taxonomy more useful.  

\subsection{Limitations}
The survey was deployed in January of 2025, prior to many of the political changes in the U.S. that may have increased concerns regarding government-involved privacy incidents. As such, these results stand as a snapshot of the incidents and harms individuals experienced at the time of the survey. Future deployments of this survey at multiple time points can provide a longitudinal perspective on how privacy incidents and harms evolve over time. 

Furthermore, while we balanced recruitment by gender and income levels, we did not recruit a nationally representative sample, nor did we recruit participants from outside the U.S. Nonetheless, our results revealed many new insights and a diversity of privacy experiences. Future work can expand to a broader sample, and to different cultures. We also encourage exploration of vulnerable and underrepresented  populations who may experience additional incidents and harms.  

This survey relies on self-report which is subject to limitations such as recall or social desirability bias. However, prior research shows that an online anonymous survey is more likely to elicit candid disclosures of sensitive privacy violations \cite{tourangeau2007sensitive}. While a survey limits our ability to follow up on participant responses, we felt it was more important to encourage participants to be open and honest about their privacy experiences. 

\subsection{Ethical Considerations}
This study is about the privacy violations and harms that individuals encounter online. The information shared is potentially sensitive and recounting those experiences could be uncomfortable. Thus, we informed participants that they could skip any questions and/or withdraw from the study at any time without waiving their compensation. At the end of the survey, participants were even given the option to exclude all their responses from analysis. No participant chose to do this. We did not collect any personally identifiable information (e.g., name, IP address) and used only pseudonymous identifiers (numeric ids between 1 and 164) before starting our analysis and curating our data. Our study protocol was reviewed and determined to be exempt by the IRB of the first author's institution.

\section{Results}
Our data consists of 369 privacy incident reports from 151 participants. In the following sections, we will discuss the types of incidents participants shared and the types of harms they encountered.   
\subsection{The Characteristics of Reported Privacy Incidents (RQ1)}

\begin{figure}[htp!]
    \centering
    \includegraphics[width=1\linewidth]{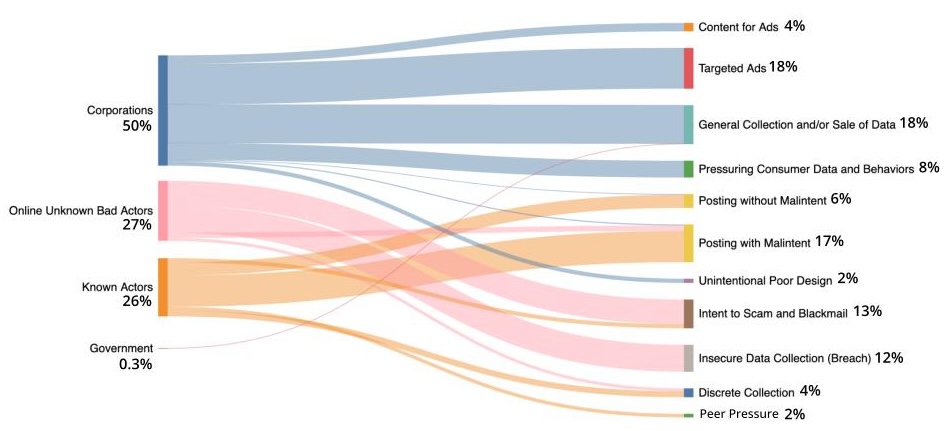}
    \caption{Mapping of actors involved in privacy incidents to the related motives. Percentage is out of total responses (N=369). Since reports can be classified in multiple categories, percentages can add up to more than 100\%.}
    \label{fig:actorsmotive}
\end{figure}

The reported incidents involved a wide range of actors with a variety of motives (see figure~\ref{fig:actorsmotive}). Additionally, many privacy experiences involved several types of information and privacy incidents (i.e., information dissemination, information collection, information processing, invasion). 


\subsubsection{The Role of Socio-Political Structures in Privacy Violations} 
\begin{figure}[htp!]
    \centering
    \includegraphics[width=1\linewidth]{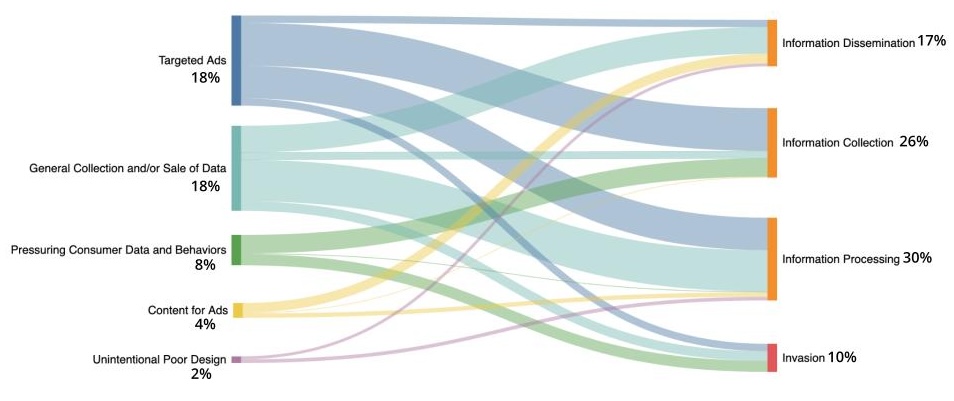}
    \caption{Mapping of motives related to socio-political structures to types of privacy incident Percentage is out of total responses (N=369)}
    \label{fig:structuresmotivseincidents}
\end{figure}

About half of the reported incidents had a socio-political structure as the main actor, with all but one involving corporations. Corporations were actors for every type of privacy incident: information dissemination, collection, processing, and invasion (see fig ~\ref{fig:structuresmotivseincidents}), and had a variety of reported motives which we describe here.

\paragraph{Targeted Ads (18\% of Total Responses)}
Targeted ads were a major motivation for corporate privacy violations, mostly (but not exclusively) involving information collection and processing violations. Participants frequently noted that their online activity was being surveilled by corporations for the purposes of targeted ads (an information collection violation), which often co-occurred with unwanted secondary data use (an information processing violation). For example, many participants noted that their browsing history was tracked and in turn they were presented targeted ads. One participant complained about, \textit{``Any Google search or Instagram ad or Facebook ad taking your search history and presenting ads to you based off of it.'' (P125)}.  

\paragraph{General Collection and/or Sale of Data (17\% of Total Responses)} Participants also shared a wide variety of incidents related to the general collection and sale of their data, which mostly (but not exclusively) involved information processing and dissemination violations. These reports predominantly included corporations surveilling participants' online activity (a collection violation), which was then sold to other parties (a dissemination violation). One participant shared, \textit{``When i looked into what information Meta was gathering from me. Social media. It made me feel like I was being watched.'' (P107}.

\paragraph{Pressuring Consumer Behaviors and Data (8\% of Total Responses)}
Some participants described how corporations interfered with their ability to make decisions regarding data disclosure. Specifically, these participants described instances where they felt forced to provide personal information in order to use a specific platform or service---which, in terms of Solove's taxonomy, is both an information collection violation and an invasion. For example, one participant shared, \textit{``I once felt pressured to share personal information while signing up for a social media platform that required detailed information, such as my phone number and date of birth, to create an account.'' (P8)}. Additionally, many of these participants voiced discomfort in not knowing why specific data was required: \textit{``When applying for things online sometimes you get asked for more information than seems required.  I'm still uneasy about sharing social security numbers and the like online.  It's become the norm however.'' (P47)}. 

\paragraph{Content for Ads (4\% of Total Responses)}
A small fraction of participants noted that they had been appropriated by corporations for use in advertisements. This involved secondary use of their online information (a processing violation). For example, one participant shared, \textit{``A paid promoter stole a photo of my wife and I'' (P115)}. This was often followed by the dissemination of their information in a way that made it falsely appear that they were endorsing topics and events: \textit{``I once found out that my photo was used in a Facebook post promoting an event I had no part in'' (P135)}. 

\paragraph{Platform Weaknesses (2\% of Total Responses)}
Finally, a few participants noted that they faced privacy violations due to unintentional design choices on the part of corporations and platforms, often leading to processing and dissemination violations. For example, one participant shared that due to a security flaw, the information they believed would stay private, was disseminated to a wide range of viewers: \textit{``It turned out that a social media platform had a security flaw that exposed parts of my private messages to unintended viewers. This happened within a messaging app, where I had assumed my conversations were secure.'' (P22)}.

\subsubsection{The Role of Individual Actors in Privacy Violations}
\begin{figure}[htp!]
    \centering
    \includegraphics[width=1\linewidth]{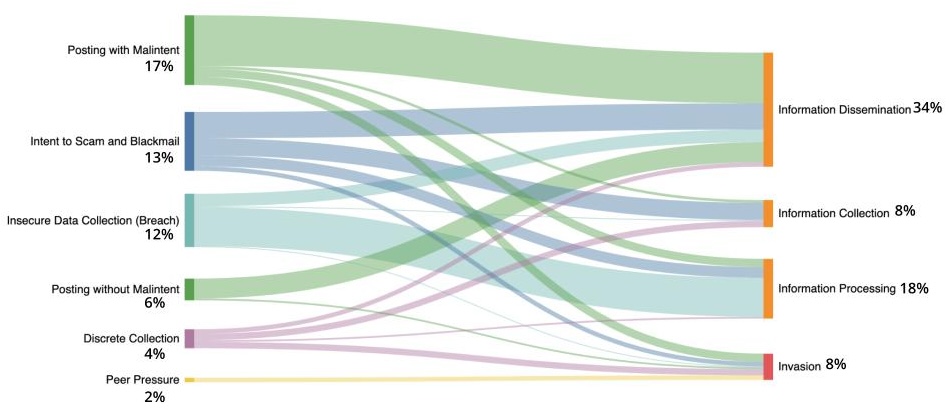}
    \caption{Mapping of motives related to individual actors to types of privacy incident. Percentage is out of total responses (N=369). As reports could be classified as multiple categories, percentages can add up to more than 100\%.}
    \label{fig:individualsmotivesincidents}
\end{figure}

The other half of the privacy violations that participants reported involved individual actors. Roughly half of these actors were people the participants personally knew (some of whom were antagonistic to the participant, and others who unintentionally caused privacy violations). The other half involved unknown actors (all of whom were noted as having malicious intent). Regardless of whether these actors were familiar, antagonistic actors tended to post about the participant with malintent; they wanted to scam and blackmail them, or initiate an unwelcome data collection (whether discretely or in an insecure way). Often, actors without ill intent accidentally subjected the participant to privacy violations (see figure~\ref{fig:actorsmotive}). These experiences predominantly (but not exclusively) included information dissemination violations (see figure ~\ref{fig:individualsmotivesincidents}).

\paragraph{Posting with Malintent (17\% of Total Responses)}
A substantial portion of participants shared experiences where people they knew offline were trying to embarrass them online by disseminating either fake or graphic information about them. Many responses noted that these violations were acts of retribution: \textit{``I had a disagreement with someone online and they blasted me on social media and told a bunch of lies to try to make me look bad to other people because I was sharing anti mlm advocacy on social media and they didn't like it.'' (P5)}. Other responses shared that these violations solely occurred in bad faith: \textit{``Naked pictures of me got leaked from someone I was talking to online.'' (P125)}.

\paragraph{Intent to Scam and Blackmail (13\% of Total Responses)}
Participants also shared experiences where online bad actors attempted to scam them and collect their personal information---these incidents involved collection, processing and dissemination violations. For example, \textit{``the person with the guise as a member of my school faculty tricked me into giving out my information which he said was going to be used for some school registrations... when they got access to my financial details siphoned money out of it and began posting information on my platform in my name'' (P18)}. Other participants shared that bad actors impersonated and disseminated fake information about them in order to attempt to scam their online connections: \textit{``I had my social media profile hacked and a separate profile was made with all of my information, this second profile was used to reach out to family and friends with someone pretending to be me.'' (P107)}.

\paragraph{Insecure Data Collection (Breach) (12\% of Responses)}
Many participant shared that their information was involved in data breaches, which are mostly processing violations. For example, one report noted, \textit{``One of my passwords and usernames was in a data breach. The message board was under a cyber attack and all of the users' information was leaked online.'' (P50)}. Additionally, some responses noted that bad actors would use the information acquired during a breach to cause further information dissemination violations: \textit{``After the ransomware attack my Reddit account has been hacked 4 times and each time they post so insanely much NSFW content that I'll never be able to remove them all one at a time(Reddit has no way to bulk remove posts you made ugh)''(P140)}.

\paragraph{Posting without Malintent (6\% of Total Responses)}
Some participants noted that their privacy was violated by personal connections disseminating their information without malintent. Some of these actors would share information about them that, while not innately sensitive or private, was still embarrassing. For example, \textit{``Friend posted a bad picture of me from a birthday which was funny until later on social media.'' (P38)}. Other actors disseminated personal information about them in an attempt to be helpful: \textit{``I had a friend post on social media a "prayer request" for me, as I was going through a rough time with depression and anxiety.'' (P145)}.

\paragraph{Discrete Collection of Data (3\% of Total Responses)}
A small fraction of participants noted that other actors attempted to secretly collect images or recordings of conversations. For example, one participant shared, \textit{``people accessing webcams when i was unaware'' (P75)}. Other participants similarly noted that their conversations were recorded, \textit{``Someone was recording a conversation I was having with them offline, still now sure why they were doing it.'' (P113)}.

\paragraph{Peer Pressure (2\% of Total Responses)}
Finally, a small fraction of participants noted that individuals they knew offline would try to pressure them to behave a certain way on social media---an invasion violation. Some noted feeling like they could not speak freely online: \textit{``While participating in a discussion forum about a sensitive topic, I felt prevented from expressing my own opinion.'' (P6)}. Others felt pressured on social media to not behave a certain way in their offline lives. For example, \textit{``I posted a picture of myself with a new tattoo on social media, specifically Instagram. I was excited about the tattoo and wanted to share it with my followers, but shortly after posting, I received several messages from friends and family, some of whom expressed concern about the design, others about the permanence of tattoos in general...made me feel pressured to reconsider my decision'' (P45)}

\subsubsection{The Types of Information Involved in Reported Incidents}

\begin{figure}[htp!]
    \centering
    \includegraphics[width=1\linewidth]{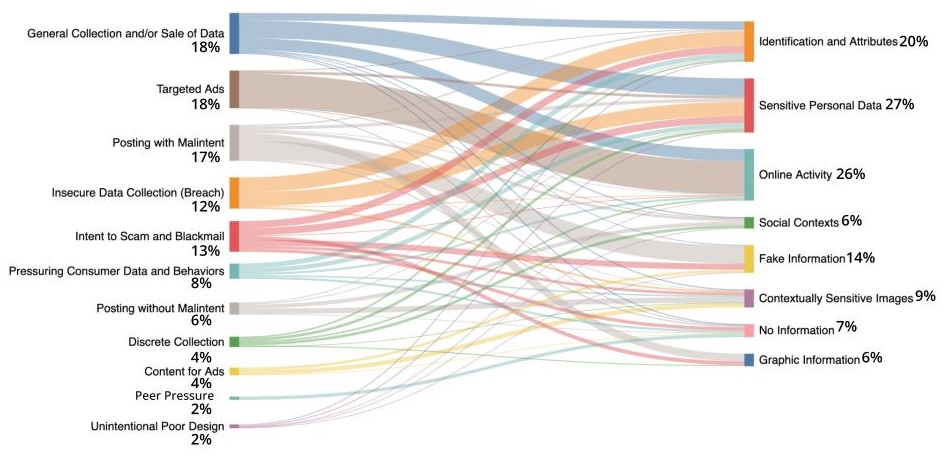}
    \caption{Mapping of underlying motive to types of information involved. Percentage is out of total responses (N=369). As reports could be classified as multiple categories, percentages can add up to more than 100\%.}
    \label{fig:motivesinfo}
\end{figure}

Across the reports, 7 types of information involved in the privacy incident were mentioned: (1) Identification and Attribute Data, (2) Sensitive Personal Data, (3) Online Activity, (4) Social Contexts, (5) Graphic Information, (6) Fake Information, (7) Contextually Sensitive Images. Table ~\ref{tab:informationtypes} displays the types of information and corresponding examples, while figure ~\ref{fig:motivesinfo} links the types of information to the motives of harm. Note that some described privacy incidents did not involve the actual collection, processing, or dissemination of information, but rather the potential for such violations to occur. 

\begin{table}[htp!]
\scriptsize
\begin{tabular}{>{\raggedright\arraybackslash}p{0.15\linewidth} >{\raggedright\arraybackslash}p{0.35\linewidth} >{\raggedright\arraybackslash}p{0.35\linewidth}}
\toprule
\textbf{Type of Information} & \textbf{Definition} & \textbf{Example} \\ \midrule
\textbf{Sensitive Personal Data} (27\% of Responses) & Data that includes financial records, health/medical information, location, and contact details. & \textit{``Someone used this information to access our network and gather personal information about myself, including \textbf{payment details}.'' (P62)} \\ \midrule
\textbf{Online Activity} (26\% of Responses) & Behavioral data of an individual's online activities (e.g., browsing history, platform usage). & \textit{``I don't not like having \textbf{ad tracking} through all of my devices.'' (P110)} \\ \midrule
\textbf{Identification and Attribute Data} (20\% of Responses) & Information which includes any form of identifying data, such as full name or Social Security Number. Additionally, it includes an individual's personal demographic attributes. & \textit{``my \textbf{social security number} was publicly posted for anyone who wanted it'' (P53)} \\ \midrule
\textbf{Fake Information} (14\% of Responses) & Fabricated information. Often in the context of rumours, lies, and impersonation. & \textit{``the kids in my highschool made fake accounts and posted on them about people even with \textbf{false information}.'' (P42)} \\ \midrule
\textbf{Contextually Sensitive Images} (9\% of Responses) & Images that are considered acceptable for certain individuals to see, but cause a violation when shared more broadly. & \textit{``it was a picture with my neighbor \textbf{normal picture}... It felt to me as unecessary as I did not see a point In that picture being posted (P33)''} \\ \midrule
\textbf{No Information Involved} (7\% of Responses) & - & \textit{``I hate when people tag me on social media without my knowledge. I've had people tag me on posts when they are trying to sell a product or tag me on scams. Its happened several times on FB and I've considered canceling all social media.'' (P86)} \\ \midrule
\textbf{Social Contexts} (6\% of Responses) & Information regarding relationship statuses and conversations had with social connections. & \textit{``Social media possibly or other websites. Websites contain a lot of my personal information as well as \textbf{connection to my family's personal information} when doing searches about myself. (P78)''} \\ \midrule
\textbf{Graphic Information} (6\% of Responses) & Any information that is graphic or explicit in nature. & \textit{``As I said before, personal \textbf{details about my sex life} linked to my full name were posted on social media after I had sex with someone else's significant other.'' (P129)} \\ \bottomrule
\end{tabular}
\caption{Codebook of types of information involved in privacy violations}
\label{tab:informationtypes}
\end{table}

\subsection{Resulting Harm (RQ2)}
\label{harmrq2}
\begin{figure}[htp!]
    \centering
    \includegraphics[width=1\linewidth]{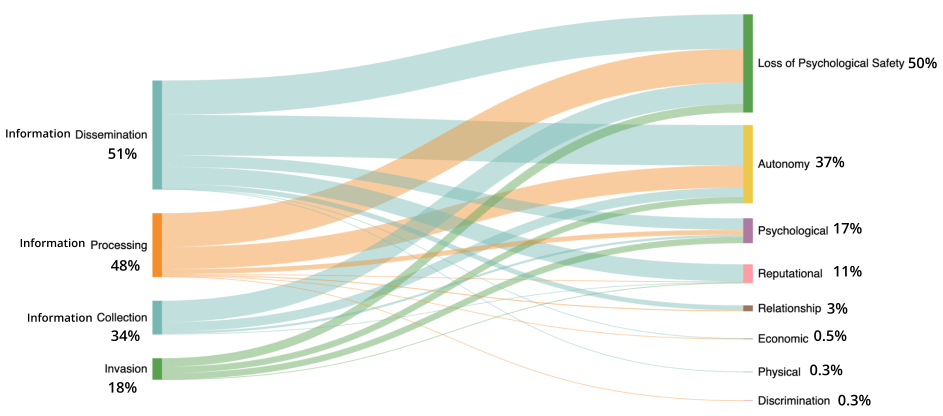}
    \caption{Mapping of reported incidents to reported harms. Percentage is out of total responses (N=369). As reports could be classified as multiple categories, percentages can add up to more than 100\%.}
    \label{fig:incidentharm}
\end{figure}

While the reported incidents were associated with a wide variety of harms (see figure~\ref{fig:incidentharm}), the predominant harms in Solove and Citron's taxonomy mentioned by participants in our study were intangible, including psychological harms, autonomy harms, reputational harms, and relationship harms. However, the most predominant type of harm was not part of the original taxonomy: loss of psychological safety. This included anxiety about future adverse events, persistent paranoia regarding technology, a general feeling of being unsafe in their daily lives, and a fear of missing out on benefits that they otherwise would have experienced.

\subsubsection{Fear of Potential Loss}
Many participants noted that due to the privacy incident they experienced, they felt like they were at a newfound risk of tangible harms, despite these harms not actually happening (yet). Participants shared concerns regarding their physical safety and their economic safety, and they were worried that some harm directly related to the incident could occur in the future. 

\paragraph{Fear of Harm Related to the Incident (11.7\% of Total Responses)} Some participants were concerned about future consequences related directly to the privacy incident. For example, one participant shared that they were concerned that someone had a recording of them that could be posted publicly, \textit{``I had a conversation of me recorded online that I did not know about. It happened on discord. I am not sure what caused it. I just did not want that conversation recorded and put out publicly.'' (P151)}. Other participants explicitly noted that a bad actor was continuously harassing and reminding them of the privacy violation and potential future harms. For instance, one participant shared, \textit{``I had an old boss who had found some old pictures of me. He used them to make me feel obligated to give more of myself to the company than should have. He had them on his phone and would flash to me every week or two just to keep me in fear.'' (P68)}. Another participant similarly shared that a stalker would continuously post photos of them online to remind them that they were being followed, \textit{``After being arrested in 2016 I had someone stalking me posting pics of me out doing random things to my fb wall. I was scared that someone was always following me and going to hurt me.'' (P156)}. 

\paragraph{Fear of Economic Harm (8.9\% of Total Responses)} Some participants reported that the unwanted online disclosure of financially-relevant information made them concerned about potential economic harm. For example, one participant shared, \textit{``Ticketmaster had a huge breach. I don't remember exactly why, but it was uncomfortable that my info (credit cards, address) were in possession.'' (P133)}. Similarly, another participant shared, \textit{``This is the same situation as described on the last page where my social security number was publicly posted for anyone who wanted it. There were probably tens of thousands in the breach. It was caused by hackers. It was a horrible situation.'' (P97)}

\paragraph{Fear of Physical Harm (5.7\% of Total Responses)} Some participants reported that the privacy incident made them concerned about potential physical harm. For instance, one participant shared that a previous stalker had found their address, leading to major concerns,\textit{ ``I had a stocker [sic] 25 years ago. I thought it was so far in the past that I was safe. However, using Facebook this man found my children (now adult) and mislead them into giving him my location and life information. This sent me back decades into fear, panic, PTSD came back in full effect.'' (P68)}. Another participant similarly shared that their address had been leaked, causing them to take extensive actions, \textit{``I have had multiple data breach incidents by major corporations where my phone number, name, address, and social security number had been shared online by what the company claims to be hacker groups. I have had to change my social security number, phone number, and move out of fear for what could happen.'' (P65)}.

\subsubsection{Persistent Paranoia}
Participants frequently expressed sentiments about feeling persistent paranoia regarding technology. Namely, a portion of participants felt like their phones were constantly listening to them, leading them to feel psychologically unsafe. Additionally, participants shared that they lost trust in technology in general. 

\paragraph{Feeling of Surveillance (13.3\% of Total Responses)} Many participants noted that they felt their phones were literally listening to them. For example, \textit{``Having conversations with others and then seeing something from that conversation online in an ad or email. That lets me know electronic devices are 'listening' and that there is none with any integrity.'' (P32)}. While these participants are likely incorrect about the exact nature of digital surveillance, their general feelings about surveillance are legitimate and should be acknowledged. Others more realistically stated that they felt constantly being tracked when using technology. For instance, one participant shared, \textit{``Whenever I get personalized ads of products or things I looked at earlier. This happens all the time when I'm looking to buy something and the next thing I see is an ad for that exact product. It occurs everywhere on my phone. I feel like it's an invasion of my privacy.'' (P114)} while another similarly shared, \textit{``well thats not so much an incident but something i always think about. whats being done with my google searches, etc. or on my iphone when an app asks to allow tracking.'' (P20)}.

\paragraph{Loss of Trust in Technologies (6.2\% of Total Responses)} Some participants noted that due to the privacy violation they incurred, they had lost trust in related technologies. For instance, one participant shared that after a contest asked for personal information, they felt, \textit{``nervous to sign up for anything anymore'' (P138)}. Another individual similarly shared that they were nervous to share information online after their username and password were involved in a data breach, \textit{``One of my passwords and usernames was in a data breach. The message board was under a cyber attack and all of the users information was leaked online. This situation made worried about sharing my information online'' (P50)}.  

\subsubsection{General Feeling of Being Unsafe}
Some participants reported a general feeling of stress, anxiety, and lack of safety in their daily lives after experiencing a privacy incident.

\paragraph{Feeling Unsafe Online (5.7\% of Total Responses)} Some participants explicitly noted feeling unsafe online due to the privacy incident. While this fear was at times due to the potential for physical harm, it also occurred without the potential for physical harm. For instance, one participant shared they were concerned because a bad actor had their address, \textit{``A man used my dating app profile pics and put me on porn. I dont know what his motive was but I even had a detective on the case because apparently he was doing this to multiple women. It caused me to be a little freaked out because this man let me know he knows where I live.'' (P152)} while another shared they felt unsafe due to their phone number being exposed, \textit{``I do not know how my information, specially personal phone number has been exposed. The incident caused unwanted contact including spam calls and massages. I consider changing my personal phone number. I felt threatening, and not safe and no privacy what so ever.'' (P67)}.

\paragraph{Feeling Stress and Anxiety (5.4\% of Total Responses)} Some participants noted feeling anxiety as they worked to mitigate their privacy violation. For instance, one participant shared, \textit{``My social media account was hacked due to a reused password, and inappropriate posts were made. It was embarrassing and stressful until I secured the account.'' (P89)}. Others noted that they faced continuous stress due to the violation: \textit{``Someone shared my phone number on social media without consent, causing unwanted messages and stress.'' (P4)}. 

\subsubsection{Missing Out}
Finally, some of the participants noted that they felt upset about missing out on things they would have experienced if a specific privacy violation had not occurred. 

\paragraph{Time and Effort Spent (5.1\% of Total Responses)} 
Some participants mentioned that they had put a considerable amount of time and effort into mitigating the harms from the violation. For example, one participant shared that after their social security information was leaked to the dark web, they spent time signing up for various identity protection programs, \textit{``I put a freeze on all three credit bureaus and signed up for identity protection program.'' (P53)}. Another participant similarly shared that after their data had been involved in many breaches, they spent considerable time changing their passwords, \textit{``I have had to change a lot of my passwords over the years. One such incident was the Playstation Network hack.'' (P160)}. Finally, some participants noted that the considerable effort they had spent curating their image online was wasted after experiencing a privacy violation. For instance, \textit{``I was online and I felt like someone was tracking my social media posts, taking them out of context and them using them to build a mischaractization of me online. This occured on a social media site and overall because i value the way i am percieved online i believe that not only did it cause damage to my image but also made me uncomfotable because it felt like the work and time of the past was wasted.'' (P13)}. 

\paragraph{Feeling Exploited (5.1\% of Total Responses)} Participants who had their images stolen for advertising purposes would often feel exploited. For instance, one participant shared, 
\textit{``I shared photos with a company and later found out that they used my photos for advertising. It was an online company where I uploaded photos from an event. I didn't want my photos displayed on their website and it made me uncomfortable that they did so without asking me.'' (P92)}. 

\subsection{Extending Solove and Citron's Privacy Taxonomies to Account for Modern Sociotechnical Systems (RQ3)}

\subsubsection{Adapted Taxonomy of Privacy Incidents}
In this section, we will be updating Solove's Taxonomy of Privacy \cite{solove2005taxonomy} based on the different experiences participants shared of privacy violations with modern-day technologies. Table ~\ref{tab:ExpandedDefinitions} displays our updated framework and the codebook of the updated framework is shown in table ~\ref{tab:IncidentCodebook}. 
 
\begin{table}[htp!]
\scriptsize
\begin{tabular}{>{\raggedright\arraybackslash}p{0.1\linewidth} | >{\raggedright\arraybackslash}p{0.1\linewidth} >{\raggedright\arraybackslash}p{0.35\linewidth} >{\raggedright\arraybackslash}p{0.35\linewidth}}
\toprule
\textbf{Category} & \textbf{Incident} & \textbf{Original Definition \cite{solove2005taxonomy}} & \textbf{Updated Definition} \\ \midrule
Information Dissemination & Distortion & Consists of the dissemination of false and misleading information about individuals & - \\ \cline{2-4}  
& Appropriation \textasciicircum{} & Involves the use of the data subject's identity to serve the aims and interests of another & Involves the use of the data subject's identity \textbf{or identifying details} to serve the aims and interests of another \\ \cline{2-4}  
& Breach of Confidentiality & Breaking a promise to keep a person's information confidential & - \\ \cline{2-4}  
& Exposure \textasciicircum{} & Involves revealing another's nudity, grief, or bodily functions & Involves revealing another's nudity, grief, bodily functions, \textbf{or contextually sensitive images} \\ \cline{2-4}  
& Spread of Personal Accessibility Information * & - & \textbf{Involves spreading or revealing another's personal contact information or temporal location} \\ \cline{2-4}  
& Disclosure & Involves the revelation of truthful information about a person that impacts the way others judge her character & - \\ \cline{2-4}  
& Blackmail & The threat to disclose personal information & - \\ \cline{2-4}  
& Increased Accessibility & Amplifying the accessibility of information & - \\ \midrule

Information Processing & Secondary Use & The use of information collected for one purpose for a different purpose without the data subject's consent & - \\ \cline{2-4}  
& Insecurity & Involves carelessness in protecting stored information from leaks and improper access & - \\ \cline{2-4}  
& Exclusion & Concerns the failure to allow the data subject to know about the data that others have about her and participate in its handling and use & - \\ \cline{2-4}  
& Aggregation & Involves the combination of various pieces of data about a person & - \\ \cline{2-4}  
& Identification \textasciicircum{} & Linking information to particular individual & Linking information to particular individual's \textbf{offline or online identity} \\ \midrule

Information Collection & Surveillance \textasciicircum{}& The watching, listening to, or recording of an individual's activities. & The watching, listening to, or recording of an individual's activites \textbf{by government, individuals, or corporations} \\ \cline{2-4}  
& Interrogation \textasciicircum{}& Consists of various forms of questioning or probing for information & Consists of various forms of questioning or probing for information \textbf{by government, individuals, or corporations} \\ \cline{2-4}  
& Deceptive Data Collection * & - & \textbf{Involves the collection of personal information by bad actors using fraudulent or deceptive mechanisms} \\ \midrule

Invasion & Intrusion & Concerns invasive acts that disturb one's tranquility or solitude & - \\ \cline{2-4}  
& Decisional Interference \textasciicircum{} & Involves the government's incursion into the data subject's decisions regarding her private affairs & Involves the government, \textbf{individuals, or corporations} incursion into the data subject's decisions regarding her \textbf{\st{private}} affairs \\ \midrule

Psychological Fear * & Deterred Bad Actors * & - & \textbf{When individuals successfully deter bad actors from using fraudulent or deceptive mechanisms against them} \\ \cline{2-4}  
& Potential for Blackmail * & - & \textbf{Consists of potentially antagonistic actors having access to information with the prospect to be used in blackmail} \\ 
\bottomrule
\end{tabular}
\caption{Definitions in Adapted Taxonomy of Privacy;  \^{} indicates extended definition; $\ast$ indicates new theme}
\label{tab:ExpandedDefinitions}
\end{table}

\begin{table}[htp!]
\scriptsize
\begin{tabular}{l | >{\raggedright\arraybackslash}p{0.2\linewidth} >{\raggedright\arraybackslash}p{0.6\linewidth}}
\toprule
\textbf{Category} & \textbf{Subcategory} & \textbf{Examples} \\ \midrule

\begin{tabular}[c]{@{}l@{}}Information \\ Dissemination \\ (49\%, N=181)\end{tabular} &
  Distortion (15\%, N=55) &
  \textit{``There was an incident a while ago where false information about me was shared on social media''} (P35) \\ \cline{2-3} 
& Appropriation (12\%, N=44) \textasciicircum{} &
  \textit{``A paid promoter stole a photo of my wife and I then shared it to their personal fb page. It made me very angry and I had to have it removed.'' (P115)} \\ \cline{2-3} 
& Confidentiality (11\%, N=39) &
  \textit{``I shared a private picture of my wedding with a close group of family and friends... I wanted to maintain a level of privacy about my personal life. However, a friend I had shared it with decided to forward the picture to a larger group of people, including some I didn't know very well.'' (P38)} \\ \cline{2-3} 
& Exposure (9\%, N=32) \textasciicircum{} &
  \textit{``When I was in high school I was in a photo on the district's main page and it wasn't that good of a photo. What caused it is they took my photo, never asked if they could use it, then put it on their website'' (P114)} \\ \cline{2-3} 
& Spread of Personal Accessibility Information (7\%, N=26) * &
  \textit{``a few years ago, a user on ebay was sharing some information about my name and address to others.'' (P36)} \\ \cline{2-3} 
& Disclosure (4\%, N=16) &
  \textit{``I had a friend post on social media a "prayer request" for me, as I was going through a rough time with depression and anxiety. I made me extremely uncomfortable as I had not asked for this, and did not care to have my mental health exposed to all of facebook.''} (P145) \\ \cline{2-3} 
& Blackmail (3\%, N=11) &
  \textit{``My computer got hacked, with spyware and was asked to pay a sum of money or my information would be deleted.'' (P21)} \\ \cline{2-3} 
& Increased Accessibility (1\%, N=5) &
  \textit{``The incident was that I was able to find my name, address, phone number and all of my immediate family member names, addresses, phone numbers shown online on a database.'' (P79)} \\ \midrule

\begin{tabular}[c]{@{}l@{}}Information \\ Processing \\ (47\%, N=173)\end{tabular} &
  Secondary Use (20\%, N=73) &
  \textit{``Shared location data for ride-hailing; app used it for targeted ads'' (P51)} \\ \cline{2-3} 
& Insecurity (16\%, N=58) &
  \textit{``The website of a public health organization was hacked and I had some of my medical records exposed.'' (P142)} \\ \cline{2-3} 
& Exclusion (11\%, N=40) &
  \textit{``This was caused by a lack of transparency in the platform's data collection and privacy policies, as well as their quest to make money by selling user data.'' (P14)} \\ \cline{2-3} 
& Aggregation (9\%, N=34) &
  \textit{``My shopping habits, location, and browsing history were clearly being tracked and shared across platforms without my consent'' (P154)} \\ \cline{2-3} 
& Identification (4\%, N=16) \textasciicircum{} &
  \textit{``I was casually browsing a shopping app, not logged into any account, thinking I was staying anonymous. But later, I received an email with personalized recommendations based on what I'd just looked at, even though I never shared any personal details.'' (P70)} \\ \midrule

\begin{tabular}[c]{@{}l@{}}Information \\ Collection \\ (32\%, N=118)\end{tabular} &
  Surveillance (19\%, N=69) \textasciicircum{} &
  \textit{``I feel like the history of my computer is out there for any company to view'' (P57)} \\ \cline{2-3} 
& Interrogation (9\%, N=33) \textasciicircum{} &
  \textit{``I once felt pressured to share personal information while signing up for a social media platform that required detailed information, such as my phone number and date of birth, to create an account.'' (P8)} \\ \cline{2-3} 
& Deceptive Data Collection (5\%, N=18) * &
  \textit{``I thought I was applying to job but it was actually a scammer trying to get my personal information'' (P39)} \\ \midrule

\begin{tabular}[c]{@{}l@{}}Invasion \\ (17\%, N=62)\end{tabular} &
  Intrusion (11\%, N=39) &
  \textit{``I was caught up in spam advertisements of betting sites'' (P63)} \\ \cline{2-3} 
& Decisional Interference (7\%, N=26) \textasciicircum{} &
  \textit{``There being a limited supply of something and the product amount decreasing in real time. It did not give me time to think if I wanted to buy it or not so I was pressured into buying it now'' (P151)} \\ \midrule

\begin{tabular}[c]{@{}l@{}}Psychological \\ Fear \\ (4\%, N=14)*\end{tabular} &
  Deterred Bad Actors (2\%, N=8) * &
  \textit{``I did not comply because the validity of the threats were not realistic, but it still made me feel incredibly vulnerable and uncomfortable.'' (P123)} \\ \cline{2-3} 
& Potential for Blackmail (2\%, N=6) * &
  \textit{{[}responding about being blackmailed{]} ``someone anonymously created a fake account on Facebook and somehow infiltrated my page, they sent all my friends and family false and disparaging information about me and my spouse.  I am not sure what triggered it'' (P48)} \\ \bottomrule

\end{tabular}
\caption{Incident Codebook; \^{} indicates extended definition; $\ast$ indicates new theme; Percentages and N's were calculated from researcher coding}
\label{tab:IncidentCodebook}
\end{table}

\paragraph{Corporations and Interpersonal Interactions Drive Most Privacy Incidents; Not the Government}
As Solove's Taxonomy of Privacy \cite{solove2005taxonomy} stems from a legal precedent, it has an emphasis on the government's role in privacy violations. For our participants though, the government was only cited once as playing a role in privacy violations. Instead, corporations were the preeminent actors causing privacy incidents. While some incident types already account for this, other incident types must be shifted from focusing fully on governmental involvement to instead focus on corporations to account for the incidents caused by modern, corporately-owned sociotechnical systems. Specifically, we extend the definitions of surveillance, decisional interference, and interrogation. 

Surveillance was initially described by Solove in the context of governmental or interpersonal actors. While interpersonal surveillance incidents did arise, governmental surveillance did not come up in any reported incidents. Instead corporate surveillance for the purpose of targeted advertising was frequently occurring. As such, we extend the definition of surveillance to be: the watching, listening to, or recording of an individual's activities \textbf{by government, individuals, or corporations}.

Similar to surveillance, decisional interference was originally defined by Solove to involve the government. Participants shared instances of interpersonal and corporate-driven decisional interference experiences as opposed to governmentally-driven violations. For instance, participants shared that due to their visibility to others on social media, their decisions were being impacted by social pressures, potentially leading them to change their behaviors. Participants also shared instances where they felt that corporations were interfering with their ability to make decisions. This often involved being required to share specific types of information, such as login, contact and financial information, in exchange for being able to use a platform or service. We note that the harms of missing out on a useful service or system may at times be considered highly detrimental, turning the demand for information in exchange for access into a decisional interference harm. As such, we extend the definition of decisional interference to be: an incident which involves the government, \textbf{individual's, or corporation's} incursion into the data subject's decisions regarding her private affairs. 

Finally, Solove's definition of interrogation does not fully align with the experiences that participants shared in their responses. Specifically, Solove discusses the legal precedent regarding interrogation from government entities, and the types of information that individuals in a position of power (i.e., employers) are allowed to ask individuals. The majority of participants that shared experiences with interrogation listed times when they were asked to provide information to platforms or corporations. In contrast with decisional interference where the pressure to disclose or make a decision is indirect and tied to some other incentive or disincentive, interrogation consists of asking the person directly for that information and pressuring them to disclose. As such, we extend the definition of interrogation to be an incident which consists of various forms of questioning or probing for information \textbf{by government, individuals, or corporations}.

\paragraph{Extended Sense of Identity}
Solove frequently described identification as connected to ``realspace'' or ``in the flesh'', meaning that data must be connected to an individual's offline identity. While many participants did share this kind of incident as an example of identification, others shared that their data was instead connected with their online identity (e.g., content shown on browser). This most frequently occurred due to the collection of data for targeted advertising. For instance, \textit{``I once searched for sensitive health-related information in a private browsing session, only to later see targeted ads about it on my social media feed. It turned out that even though I [wasn't] logged in, my activity was still being tracked and linked to my profile through device fingerprints and hidden trackers.'' (P154)}. As such, we expand the definition of identification to be: linking information to particular individual's \textbf{offline or online identity}. 

Participants reported experiencing appropriation differently than defined by Solove. While Solove defines appropriation as an individual having their likeness or personality used for another individuals' gain, our participants more frequently cited situations where their personal contact details were appropriated: \textit{``Somehow scam callers always call me about my cars extended warranty or how I no longer have health insurance and they know my name, the city I live in, and obviously have my phone number.'' (P114)}. This indicates that people may have an extended view of what their likeness and personality includes. As such, we expand the definition of appropriation to be: an incident that involves the use of the data subject's identity \textbf{or identifying details} to serve the aims and interests of another.   

\paragraph{The Type of Information Involved has Expanded}
Whereas Solove focuses on very specific data types to define privacy, we notice a shift to a broader set of contextually sensitive information. Namely, many participants shared that they experienced exposure with information that wasn't necessarily as explicit as described by Solove. Instead, participants felt exposed by information that was embarrassing, but not necessarily graphic or sensitive, or by information that they allowed to be shared with one audience but not another. As such, we extend the definition of exposure to be: an incident that involves revealing another's nudity, grief, bodily functions, \textbf{or contextually sensitive images}. 

Similarly, Solove describes decisional interference incidents as involving interference with ``home, family, and body''\cite{solove2005taxonomy}. Across our participants' responses we saw that decisional interference effected a broader range of categories. As such, we further update the definition of decisional interference to be: decisional interference involves the government, individuals, or corporations incursion into the data subject's decisions regarding her \textbf{\st{private}} affairs. 

Finally, we found that many participants described instances where their private contact information and temporal location were disseminated to a variety of audiences without their consent. A breach of confidentiality was often not associated with this, and this type of information also did not fall under the definition of increased accessibility as the information was not initially public. Additionally this type of incident is also not considered disclosure, as the dislcosure of the information would not inherently change the way others perceive the participant's character. Currently, in Solove's framework, this type of dissemination is not explained.  As such, we add a new incident type to Solove's Taxonomy: \textit{Spread of Personal Accessibility Information}. This incident type \textbf{involves spreading or revealing another's personal contact information or temporal location}.

\paragraph{Deceptive Data Collection}
6.8\% of total responses in our study shared experiences with attempted scams online. While a portion of these responses fit under Solove's definition of blackmail, many did not. Instead, the responses discussed how bad actors were using deceptive data collection techniques to collect private information. For instance, bad actors would impersonate social media users to collect data from friends and family. As the use of deception from bad actors to collect private information is currently unaccounted for in Solove's Taxonomy of Privacy, we introduce a new incident type, \textit{Deceptive Data Collection}, which is defined as \textbf{involving the collection of personal information by bad actors using fraudulent or deceptive mechanisms}.

\paragraph{New Found Psychological Fears}
There were 14 participants who shared 'Psychological Fear' incidents (4\% of total responses): they felt that there was a potential for an incident to occur, despite it not actually occurring. 
For instance, a portion of participants who responded that they had been blackmailed, shared that they \textit{felt} like they were being blackmailed because other individuals had access to potentially damaging information about them. These participants never noted that they were asked to provide something in exchange for non-disclosure, though. For instance, one participant shared that someone had access to a private photo that they had sent them, \textit{``I had sold a picture of myself (no face) to someone and they posted it to Reddit in connection with my Tik Tok username.  It made me feel disgusting and ashamed, and I had to report it to get it taken down.'' (P87)}. Another participant similarly shared, \textit{``Private photos taken by an ex friend. We went separate ways. On personal cellphone camera. Was afraid/ashamed for the pictures to be showed or shared with others'' (P99)}. To account for this type of incident, we include \textit{Potential for Blackmail} in our updated framework. This is defined as: an incident that \textbf{consists of potentially antagonistic actors having access to information with the prospect to be used in blackmail}.

Similarly, some participants noted incidents where bad actors attempted to scam them, but were unsuccessful: \textit{``It was a scam email from someone I've never heard of. They tried to extort me for crypto under the threat that they had compromising photos of me. I know no such photos exist, and my devices always have the camera lense covered when not in use specifically to prevent them from existing...'' (P97)}. While the perpetrators were unsuccessful in scamming the participant, they still shared that they felt harmed as a result, \textit{``...Even though I am generally aware of this kind of thing and knew it was an extortion scam, I still felt uneasy knowing how many people that have shared compromising photographs would feel in that situation.'' (P97)}. To account for this type of incident, we include, \textit{attempted scam} in our updated framework, which is defined as: \textbf{when individuals successfully deter bad actors from using fraudulent or deceptive mechanisms against them}.

\subsubsection{Adapted Typology of Privacy Harms}

\begin{table}[htp!]
\scriptsize
\begin{tabular}{>{\raggedright\arraybackslash}p{0.15\linewidth} | >{\raggedright\arraybackslash}p{0.15\linewidth} >{\raggedright\arraybackslash}p{0.6\linewidth}}
\toprule
\textbf{Category} & \textbf{Harm} & \textbf{Example} \\ \midrule

\begin{tabular}[c]{@{}l@{}}Loss of Psychological \\ Safety * \\ (50\%, N=185)\end{tabular} &
  Fear of Potential Loss* (20\%, N=73) &
  \textit{“I have had multiple data breach incidents by major corporations where my phone number, name, address, and social security number had been shared online by what the company claims to be hacker groups. I have had to change my social security number, phone number, and move out of fear for what could happen.” (P65)} \\ \cline{2-3} 
& Persistent Paranoia* (19\%, N=69) &
  \textit{“Having conversations with others and then seeing something from that conversation online in an ad or email. That lets me know electronic devices are 'listening' and that there is none with any integrity. By none I mean government and business of any kind.” (P32)} \\ \cline{2-3} 
& General Feeling of Being Unsafe* (10\%, N=38) &
  \textit{“I do not know how my information, specially personal phone number has been exposed. The incident caused unwanted contact including spam calls and massages. I consider changing my personal phone number. I felt threatening, and not safe and no privacy what so ever.” (P67)} \\ \cline{2-3} 
& Missing Out* (10\%, N=36) &
  \textit{“Once again, just feeling exploited on the Internet and feeling like people have access to my information on my ad preferences” (P46)} \\ \midrule

\begin{tabular}[c]{@{}l@{}}Autonomy Harms \\ (37\%, N=138)\end{tabular} &
  Lack of Control (25\%, N=92) &
  \textit{``The incident occurred when trying to cancel a subscription. The incident was caused by me trying to change a subscription to a cheaper price. The incident occurred on a company website. The incident caused me discomfort because I felt like I was forced to keep the plan I have.'' (P1)} \\ \cline{2-3} 
& Thwarted Expectations (20\%, N=73) &
  \textit{``I signed up to learn more about an affiliate marketing program and afterwards I was bombarded with emails to the email that was used to sign up.'' (P92)} \\ \cline{2-3} 
& Failure to Inform (10\%, N=37) &
  \textit{``Several times companies do not tell you what they use your information for.'' (P110)} \\ \cline{2-3} 
& Coercion (7\%, N=25) &
  \textit{“I had a crazy ex girlfriend. She had possession of some photos I didn't want out there. She was demanding a payment for them.” (P115)} \\ \cline{2-3} 
& Chilling Effect (4\%, N=15) &
  \textit{“I made a post on facebook on a controversial topic. I did not post anything offensive but merely explaining my stance. Unfortunately an ex friend let everyone know who originally posted it…I basically stopped posting on Facebook and to this day I barely do.” (P103)} \\ \cline{2-3} 
& Manipulation (3\%, N=11) &
  \textit{“On the tiktok social media app I felt pressure to opt into allowing them to track my online activity to better their algorithm. They made it seem like the app would be less useful if I did not do this. It made me feel uncomfortable seeing ads for things that they knew I was currently researching.” (P50)} \\ \midrule

\begin{tabular}[c]{@{}l@{}}Psychological Harms \\ (17\%, N=64)\end{tabular} &
  Disturbance (11\%, N=39) &
  \textit{“People kept messaging me on my work email late after the work day had ended.” (P73)} \\ \cline{2-3} 
& Emotional Distress (8\%, N=30) &
  \textit{“One time I told someone where I used to work on a social media app and they shared it with their friends. They started tagging me and making fun of my employer. It was infuriating because I’m a very private person.” (P160)} \\ \midrule

\begin{tabular}[c]{@{}l@{}}Reputational Harms \\ (11\%, N=42)\end{tabular} &
  \textbf{-} &
  \textit{``It harmed my reputation... and caused people to view me differently for no reason!'' (P126)} \\ \midrule

\begin{tabular}[c]{@{}l@{}}Relationship Harms \\ (3\%, N=12)\end{tabular} &
  \textbf{-} &
  \textit{“someone anonymously created a fake account on Facebook and somehow infiltrated my page, they sent all my friends and family false and disparaging information about me and my spouse. I am not sure what triggered it by I think it was his ex. This caused me a lot of mental issues and almost destroyed my marriage.” (P48)} \\ \midrule

\begin{tabular}[c]{@{}l@{}}Economic Harms \\ (0.5\%, N=2)\end{tabular} &
  \textbf{-} &
  \textit{“a major company's information was hacked so they got my name, address, etc. The violator then tried to open credit cards in my name with charges of large amounts.” (P57)} \\ \midrule

\begin{tabular}[c]{@{}l@{}}Physical Harms \\ (0.3\%, N=1)\end{tabular} &
  \textbf{-} &
  \textit{“I had someone that I had a disagreement with spread rumors online about me, including my personal details, address, phone, and more…  I received numerous messages about the claims, including publicly in comments, as well as phone calls, text messages, and even a rock thrown at my door.” (P65)} \\ \midrule

\begin{tabular}[c]{@{}l@{}}Discrimination Harms \\ (0.3\%, N=1)\end{tabular} &
  \textbf{-} &
  \textit{“I was giving information for a bank loan on a website and while I was giving them background information that shouldnt have been relevant they were altering the amount I could recieved depending on those factors even though they said these wouldnt be a part of the overall application.” (P13)} \\ \bottomrule

\end{tabular}
\caption{Adapted Typology of Harms; * indicates new code}
\label{tab:HarmsCodebook}
\end{table}

In section ~\ref{harmrq2} we already described the types of harms unaccounted for in Solove and Citron's Typology of Harm. To summarize, one of the main findings was that harms traditionally found in a tort law context, such as economic, physical and discrimination harms, were rarely reported. Instead, our dataset contained a high frequency of harms regarding autonomy, psychological well-being, and concerns regarding loss of psychological safety. Psychological harms defined in Solove and Citron's typology did not cover the complexity of experiences participants shared regarding these types of harm. Namely, many participants shared having an overall loss of psychological safety, or feeling concerns regarding future issues that could arise from the privacy violation. As such, we expand Solove and Citron's framework to include the category \textit{Loss of Psychological Safety}, which has 4 relevant subcategories based on our participant responses:
\begin{itemize}
    \item \textbf{Fear of Potential Loss:} The fear of potential tangible privacy harms due directly to a privacy violation
    \item \textbf{Persistent Paranoia:} Increased paranoia of modern technologies after a privacy violation
    \item \textbf{General Feeling of Being Unsafe:} A lack of perceived general safety after a privacy violation.
    \item \textbf{Missing Out:} Concern regarding the potential of missing out on future personal gain.
\end{itemize}

We found that the way participants experienced existing harms (autonomy harms, reputational harms, relationship harms, psychological harms, economic harms, and physical harms) aligned with how they were described by Solove and Citron. These therefore remain unaltered in our updated taxonomy. See table ~\ref{tab:HarmsCodebook} for the codebook of the expanded framework of harms. Refer back to section ~\ref{harmrq2} for extended details.

\section{Discussion}
While we were able to largely align our survey data with Solove's taxonomies, we note several salient points of departure. For one, Solove's Taxonomy of Privacy \cite{solove2005taxonomy} encapsulates existing legal precedent at the time of its publication. As such, many of the incident types in the framework are concerned with privacy incidents that arise from specific contexts or particular actors that were involved in legal actions. In contrast, our results covered a broader range of contexts and actors that are involved in privacy incidents. Furthermore, whereas Solove and Citron's Typology of Privacy Harms mostly covers harms that can be legally defined, participants in our sample frequently noted a loss of psychological safety (e.g., feeling severe concerns regarding potential future tangible harms) as a legitimate harm. This indicates that even though severe physical or economic harms infrequently occur, individuals are acutely aware of them, which makes them feel anxious and distressed. Our updated taxonomy captures this stress and anxiety.

\subsection{Implications for Research}
As privacy is multi-faceted \cite{knijnenburg2022modern}, we need to understand a variety of its components. By adapting Solove's Taxonomy of Privacy \cite{solove2005taxonomy} with a more human-centered approach, we present a more general framework that can be more readily applied across the field of HCI. Prior work \cite{lee2024deepfakes} has allowed for researchers to understand the technical components of how AI changes the \textit{mechanisms} of the privacy incidents in Solove's Taxonomy of Privacy \cite{solove2005taxonomy}. Our work uncovers how individuals experience or \textit{perceive} these types of incidents. This is important information for researchers to understand, as perceptions have been shown to lead to changes in behavior \cite{fishbein1975belief}. Understanding both how technologies exacerbate privacy incidents and how individuals perceive such incidents can be beneficial in better understanding the multi-faceted nature of privacy. Therefore, our updated taxonomy provides a potential point of integration with other work that has examined Solove's framework with specific technologies, such as AI.

Additionally, our analysis resulted in relatively fewer changes to Solove and Citron's Typology of Privacy Harms \cite{citron2022privacy}. While there were some additional harms that were unaccounted for in the taxonomy, the originally specified harms still occurred. Our main departure in the updated taxonomy is that the \textit{fear} of harms were more common than the harms themselves. For instance, while there were not many reported physical harms, the fear of physical harm was prevalent among participants. Despite the stability of the harms framework, the incidents which led to these harms changed considerably from Solove's initial taxonomy \cite{solove2005taxonomy}, which calls its use in contemporary HCI research into question. Specifically, in adopting conceptual frameworks, privacy scholars should be more aware that the mechanisms we are studying are changing, and that these changes either require a continually updated framework, or a more stable conceptualization of the work. Specifically, given the relative stability of the harms framework, researchers should consider conceptualizing their work from the perspectives of harms, and to take a more open-ended approach as to why the harms occurred. This could involve asking research participants about harms rather than about incidents, or, conduct interviews or open-ended surveys when aiming to capture incidents. This could lead to discovering new ways that harms are occurring.

\subsection{Implications for Policy}
Solove's Taxonomy of Privacy \cite{solove2005taxonomy} is based on existing legal precedent which focuses on specific actors, underlying causes, and types of information involved. In studying people's lived privacy incidents and harms, we found the framework had to be expanded to allow for a broader variety of actors, causes, and information. This indicates that there may be holes in legislation due to unaccounted for actors, unregulated causes, and novel types of information that require special-purpose legislation. For example, the European Union's General Data Protection Regulation (GDPR) is used to widely legislate against privacy violations. However, this regulation does not apply to ``the processing of personal data by a natural person in the course of a purely personal or household activity and thus with no connection to a professional or commercial activity. Personal or household activities could include correspondence and the holding of addresses, or social networking and online activity undertaken within the context of such activities. However, this Regulation applies to controllers or processors which provide the means for processing personal data for such personal or household activities. (Recital 18)'' \cite{GDPR}. Many of the privacy incidents we saw, such as secondary use violations, were enacted by individuals who had no connection to professional or commercial activity, therefore they are not covered in this regulation. New policy and frameworks should consider the broadening of actors and motives behind the incidents that individuals are experiencing.

Our findings also uncover the importance of the temporal dimension of privacy: individuals are anticipating future harms, leading to long-term pervasive fear among individuals, which then becomes a harm in its own right. Legally, we are increasingly seeing this anticipated harm reflected in information risk case precedent: A paper trail can be used to show misconduct and incriminate. However, the act of not properly retaining information to see if misconduct has occurred is problematic. This was seen in infamous cases like Arthur Anderson and Enron (paper shredding cases) \cite{li2010case} where courts decided that companies that did not have a process in place to properly archive and delete documents could be punished regardless of whether misconduct had actually occurred. The parallel is that regardless of whether an actor is harming individuals financially or physically, if there is no mechanism in place to prevent such harms, this will raise concerns. Legally, these concerns could be accounted for by a court-admission of arguments around negligence in privacy-related lawsuits.

\section{Conclusion}
 We ran an empirical survey study to understand lived privacy incidents and harms, and found that there were a variety of factors, such as a broader scope of actors and information types, unaccounted for in existing privacy frameworks based on legal precedent. Additionally, we found that while many individuals have not experienced tangible harms, they instead face persistent concern over the potential for said harms. Based on our findings, we updated two existing taxonomies of privacy incidents and privacy harms. These updated taxonomies can be used by privacy researchers to situate and contextualize their work.

\bibliographystyle{ACM-Reference-Format}
\bibliography{bibliography}

\appendix
\section{Existing Privacy Frameworks}
\label{appendixFrameworks}
Solove's Taxonomy of Privacy \cite{solove2005taxonomy} is shown on table ~\ref{tab:taxofprivacy}. Citron and Solove's Typology of Harms \cite{citron2022privacy} is shown on table ~\ref{tab:harmstaxonomy}.

\begin{table}[H]
\scriptsize
\begin{tabular}{l | >{\raggedright\arraybackslash}p{0.23\linewidth} >{\raggedright\arraybackslash}p{0.55\linewidth}}
\toprule
\textbf{Category} & \textbf{Incident} & \textbf{Definition \cite{solove2005taxonomy}} \\ \midrule

\begin{tabular}[c]{@{}l@{}}Information \\ Dissemination\end{tabular} & Breach of Confidentiality & Breaking a promise to keep a person's information confidential \\ \cline{2-3} 
& Distortion & Consists of the dissemination of false and misleading information about individuals \\ \cline{2-3} 
& Exposure & Involves revealing another's nudity, grief, or bodily functions \\ \cline{2-3} 
& Disclosure & Involves the revelation of truthful information about a person that impacts the way others judge her character \\ \cline{2-3} 
& Appropriation & Involves the use of the data subject's identity to serve the aims and interests of another \\ \cline{2-3} 
& Blackmail & The threat to disclose personal information \\ \cline{2-3} 
& Increased Accessibility & Amplifying the accessibility of information \\ \midrule

\begin{tabular}[c]{@{}l@{}}Information \\ Processing\end{tabular} & Insecurity & Involves carelessness in protecting stored information from leaks and improper access \\ \cline{2-3} 
& Aggregation & Involves the combination of various pieces of data about a person \\ \cline{2-3} 
& Secondary Use & The use of information collected for one purpose for a different purpose without the data subject's consent \\ \cline{2-3} 
& Exclusion & Concerns the failure to allow the data subject to know about the data that others have about her and participate in its handling and use \\ \cline{2-3} 
& Identification & Linking information to a particular individual \\ \midrule

\begin{tabular}[c]{@{}l@{}}Information \\ Collection\end{tabular} & Surveillance & The watching, listening to, or recording of an individual's activities \\ \cline{2-3} 
& Interrogation & Consists of various forms of questioning or probing for information \\ \midrule

Invasion & Intrusion & Concerns invasive acts that disturb one's tranquility or solitude \\ \cline{2-3} 
& Decisional Interference & Involves the government's incursion into the data subject's decisions regarding her private affairs \\ \bottomrule

\end{tabular}
\caption{Solove's Taxonomy of Privacy}
\label{tab:taxofprivacy}
\end{table}

\begin{table}[H]
\scriptsize
\begin{tabular}{>{\raggedright\arraybackslash}p{0.15\linewidth} | >{\raggedright\arraybackslash}p{0.15\linewidth} >{\raggedright\arraybackslash}p{0.55\linewidth}}
\toprule
\textbf{Category} & \textbf{Subcategory} & \textbf{Definitions \cite{citron2022privacy}} \\ \midrule

\textbf{Autonomy Harms} & Lack of Control & Involves the inability to make certain choices about one's personal data or to be able to curtail certain uses of the data \\ \cline{2-3} 
& Thwarted Expectations & Involves the undermining of people's choices, such as breaking promises made about the collection, use, and disclosure of personal data \\ \cline{2-3} 
& Failure to Inform & Involves failing to provide individuals with information to assist them in making informed choices about their personal data or exercise of their privacy rights \\ \cline{2-3} 
& Coercion & Involves a constraint or undue pressure on one's freedom to act or choose \\ \cline{2-3} 
& Chilling Effect & Involves harm caused by inhibiting people from engaging in certain civil liberties, such as free speech, political participation, religious activity, free association, freedom of belief, and freedom to explore ideas \\ \cline{2-3} 
& Manipulation & Involves undue influence over a person's behavior or decision-making \\ \midrule

\textbf{Psychological Harms} & Disturbance & Involves unwanted intrusions that disturb tranquility, interrupt activities, sap time, and otherwise serve as a nuisance \\ \cline{2-3} 
& Emotional Distress & Encompasses a wide range of emotions including annoyance, frustration, fear, embarassment, anger, and various degrees of anxiety \\ \midrule

\textbf{Reputational Harms} & \textbf{-} & Involve injuries to an individual's reputation and standing in the community \\ \midrule
\textbf{Relationship Harms} & \textbf{-} & Involve the damage to relationships that are important for one's health, well-being, life activities, and functioning in society \\ \midrule
\textbf{Economic Harms} & \textbf{-} & Involve monetary losses or a loss in the value of something \\ \midrule
\textbf{Physical Harms} & \textbf{-} & Harms that result in bodily injury or death \\ \midrule
\textbf{Discrimination Harms} & \textbf{-} & Involve entrenching inequality and disadvantaging people based on gender, race, national origin, sexual orientation, age, group membership, or other characteristics or affiliations \\ \bottomrule

\end{tabular}
\caption{Solove and Citron's Privacy Harms}
\label{tab:harmstaxonomy}
\end{table}

\section{Survey Instrument}
\label{survey}
\subsection{Close-Ended Questions about Privacy Incident}
An \textbf{online privacy incident} is when an individual or organization violates your expectations of how your information should be used or how they should interact with you online. Think back to your experiences online. Below is a series of privacy incidents, please read about each type and then answer if you have experienced a similar privacy incident.

\begin{enumerate}
    \item Have you ever felt that someone or an organization was closely watching or tracking your actions online in a way that made you feel uncomfortable?
    \begin{itemize}
        \item Yes
        \item No
        \item I'm not sure
    \end{itemize}
    \item Have you ever felt pressured or forced to share personal information online?
    \begin{itemize}
        \item Yes
        \item No
        \item I'm not sure
    \end{itemize}
    \item Have you ever felt pressured or prevented from making your own decisions online?
    \begin{itemize}
        \item Yes
        \item No
        \item I'm not sure
    \end{itemize}
    \item Have you ever had your name, image, likeness, or other personal details used without your permission to promote something or for someone else’s gain in a way that made you feel uncomfortable?
    \begin{itemize}
        \item Yes
        \item No
        \item I'm not sure
    \end{itemize}
    \item Have you ever noticed that different pieces of your information (e.g., shopping habits, location, or social media activity) were combined in a way that made you feel uncomfortable?
    \begin{itemize}
        \item Yes
        \item No
        \item I'm not sure
    \end{itemize}
    \item Have you ever had your anonymous data or activity linked back to your identity in a way that made you feel uncomfortable?
    \begin{itemize}
        \item Yes
        \item No
        \item I'm not sure
    \end{itemize}
    \item Have you ever had your personal information exposed or misused because it wasn’t properly protected?
    \begin{itemize}
        \item Yes
        \item No
        \item I'm not sure
    \end{itemize}
    \item Have you ever shared information with someone or an organization and then they used it for a different purpose in a way that made you feel uncomfortable?
    \begin{itemize}
        \item Yes
        \item No
        \item I'm not sure
    \end{itemize}
    \item Have you ever been kept in the dark about what type of personal information was being used by someone or an organization, or how it was used?
    \begin{itemize}
        \item Yes
        \item No
        \item I'm not sure
    \end{itemize}
    \item Have you ever shared private information online with someone or an organization only to find out they shared it with someone else without your permission?
    \begin{itemize}
        \item Yes
        \item No
        \item I'm not sure
    \end{itemize}
    \item (Attention Check) For this question please select I'm not sure.
    \begin{itemize}
        \item Yes
        \item No
        \item I'm not sure
    \end{itemize}
    \item Have you ever had personal information about you shared publicly in a way that made you feel uncomfortable?
    \begin{itemize}
        \item Yes
        \item No
        \item I'm not sure
    \end{itemize}
    \item Have you ever had embarrassing or sensitive details about you shared publicly (e.g., photos or private moments shared without your consent) in a way that made you feel uncomfortable?
    \begin{itemize}
        \item Yes
        \item No
        \item I'm not sure
    \end{itemize}
    \item Have you ever had information about you made much more accessible for others or organizations (e.g., your address or financial details) in a way that made you feel uncomfortable?
    \begin{itemize}
        \item Yes
        \item No
        \item I'm not sure
    \end{itemize}
    \item (Attention Check) You have never had your anonymous data or activity linked back to your identity in a way that made you feel uncomfortable
    \begin{itemize}
        \item Yes
        \item No
        \item I'm not sure
    \end{itemize}
    \item Have you ever been threatened with the release of personal or private information unless you met someone’s demands?
    \begin{itemize}
        \item Yes
        \item No
        \item I'm not sure
    \end{itemize}
    \item Have you ever had false or misleading information about you shared in a way that damaged your reputation or how others see you?
    \begin{itemize}
        \item Yes
        \item No
        \item I'm not sure
    \end{itemize}
    \item Have you ever felt invaded in your online personal space, or during private moments offline?
    \begin{itemize}
        \item Yes
        \item No
        \item I'm not sure
    \end{itemize}

\end{enumerate}

\subsection{Open-Ended Questions about Three Randomly Assigned Incidents}
\textit{Participants answered this set of questions 3 times and were assigned 3 of the following conditions}
\begin{itemize}
    \item you felt that someone or an organization was closely watching or tracking your actions online in a way that made you feel uncomfortable
    \item you felt pressured or forced to share personal information online
    \item you noticed that different pieces of your information (e.g., shopping habits, location, or social media activity) were combined in a way that made you feel uncomfortable
    \item you had your anonymous data or activity linked back to your identity in a way that made you feel uncomfortable
    \item you had your personal information exposed or misused because it wasn't properly protected
    \item you shared information with someone or an organization and then they used it for a different purpose in a way that made you feel uncomfortable
    \item you felt you were kept in the dark about what type of personal information was being used by someone or an organization, or how it was used
    \item you shared private information online with someone or an organization only to find out they shared it with someone else without your permission
    \item you had personal or sensitive information about you shared publicly without your consent
    \item you had had embarrassing or sensitive details about you shared publicly (e.g., photos or private moments shared without your consent) in a way that made you feel uncomfortable
    \item you had information about you made much more accessible for others or organizations (e.g., your address or financial details) in a way that made you feel uncomfortable
    \item you were threatened with the release of personal or private information unless you met someone’s demands
    \item you had your name, image, likeness, or other personal details used without your permission to promote something or for someone else's gain in a way that made you feel uncomfortable
    \item you had false or misleading information about you shared in a way that damaged your reputation or how others see you
    \item you felt invaded in your online personal space, or during private moments offline
    \item you felt pressured or prevented from making your own decisions online
\end{itemize}

\begin{enumerate}
    \item Please describe the situation where \underline{\hspace{3cm}}. Pleaes include the following (Open-ended):
    \begin{enumerate}
        \item What was the incident
        \item What caused the incident
        \item Where the incident occurred (social media, mobile app, etc)
        \item How the incident caused you discomfort or harm
    \end{enumerate}
\end{enumerate}

\subsection{Demographics}
\begin{enumerate}
    \item (Quality Check) The fence is red. What color is the fence? (Open-Ended)
    \item \textbf{This answer will not affect your compensation for participating in the study.} We realize that sometimes chatGPT and other AI-powered tools can help with editing and grammar, but it is important for us to understand when they are used to edit responses. Did you use any of these tools to help you generate answers to the survey? This question will not affect your compensation. (Open-Ended)
    \item \textbf{If you answered yes,} please explain how you used chatGPT or other AI-powered tools to generate your answers. (Open-Ended)
    \item Do you have any other comments or questions related to this study? (Open-Ended)\\
    \noindent Finally, we like to understand the different people participating in our study. Please tell us about yourself.
    \item What sex were you assigned at birth, on your original birth certificate?
    \begin{itemize}
        \item Female
        \item Male
        \item Other (write-in)
        \item Prefer not to say
    \end{itemize}
    \item How do you currently describe yourself (mark all that apply)?
    \begin{itemize}
        \item Female
        \item Male
        \item Transgender
        \item Prefer not to say
        \item I use a different term (write-in)
    \end{itemize}
    \item Which of the following do you consider yourself to be? Select as many as apply.
    \begin{itemize}
        \item Gay or Lesbian
        \item Straight, that is not gay or lesbian
        \item Bisexual
        \item Prefer not to say
        \item I don't know
        \item I use a different term (write-in)
    \end{itemize}
    \item What is your age?
    \begin{itemize}
        \item Under 18
        \item 18 - 24
        \item 25 - 34
        \item 35 - 44
        \item 45 - 54
        \item 55 -64
        \item 65+
        \item Prefer not to say
    \end{itemize}
    \item Please indicate whether or not you have ever used the following social media sites (Yes/No).
    \begin{itemize}
        \item Instagram
        \item X, formerly known as Twitter
        \item LinkedIn
        \item VirtualLink
        \item Facebook
        \item Snapchat
        \item TikTok
    \end{itemize}
    \item Are you Hispanic or Latino?
    \begin{itemize}
        \item Yes, Hispanic or Latino
        \item No, not Hispanic or Latino
        \item Prefer not to say
    \end{itemize}
    \item What is your race? Select all that apply.
    \begin{itemize}
        \item American Indian or Alaska Native
        \item Asian
        \item Black or African American
        \item Hispanic or Latino
        \item Native Hawaiian or other Pacific Islander
        \item White
        \item Other (Please Specify)
        \item Prefer not to say
    \end{itemize}
    \item What is your highest level of education?
    \begin{itemize}
        \item Some secondary education
        \item Graduated secondary education
        \item Some college/university
        \item Graduated college/university
        \item Started post-graduate degree
        \item Post-graduate degree (Masters, PhD, JD, etc.)
        \item Prefer not to say
    \end{itemize}
    \item (Quality Check) To make sure that you aren't a bot, carrots are:
    \begin{itemize}
        \item Rocks
        \item Vegetables
        \item Fruits
        \item Animals
    \end{itemize}
    \item What is your household income level? 
    \begin{itemize}
        \item Less than \$15,000
        \item \$15,000 to \$34,999
        \item \$35,000 to \$49,999
        \item \$50,000 to \$74,999
        \item \$75,000 to \$99,999
        \item \$100,000 or more
        \item Prefer not to say
    \end{itemize}
    \item What is your present religion, if any? 
    \begin{itemize}
        \item Agnostic
        \item Atheist
        \item Buddhist
        \item Hindu
        \item Jehovah's Witness
        \item Jewish
        \item Mormon (Church of Jesus Christ of Latter-day Saints or LDS)
        \item Muslim
        \item Protestant (for example, Baptist, Methodist, Nondenominational, Lutheran, Presbyterian, Pentecostal, Episcopalian, Reformed, Church of Christ, etc.)
        \item Roman Catholic
        \item Something else, please specify
        \item Nothing in particular
        \item Prefer not to say
    \end{itemize}
    \item Aside from Weddings and funerals, how often do you attend religious services?
    \begin{itemize}
        \item More than once a week
        \item Once a week
        \item Once or twice a month
        \item A few times a year
        \item Seldom
        \item Never
        \item Prefer not to say
    \end{itemize}
    \item Do you identify as a person with a disability or other chronic conditions?
    \begin{itemize}
        \item Yes
        \item No
        \item Prefer not to say
    \end{itemize}
\end{enumerate}

\section{Full SanKey Diagrams}
\begin{figure}[htp!]
    \centering
    \includegraphics[width=0.75\linewidth]{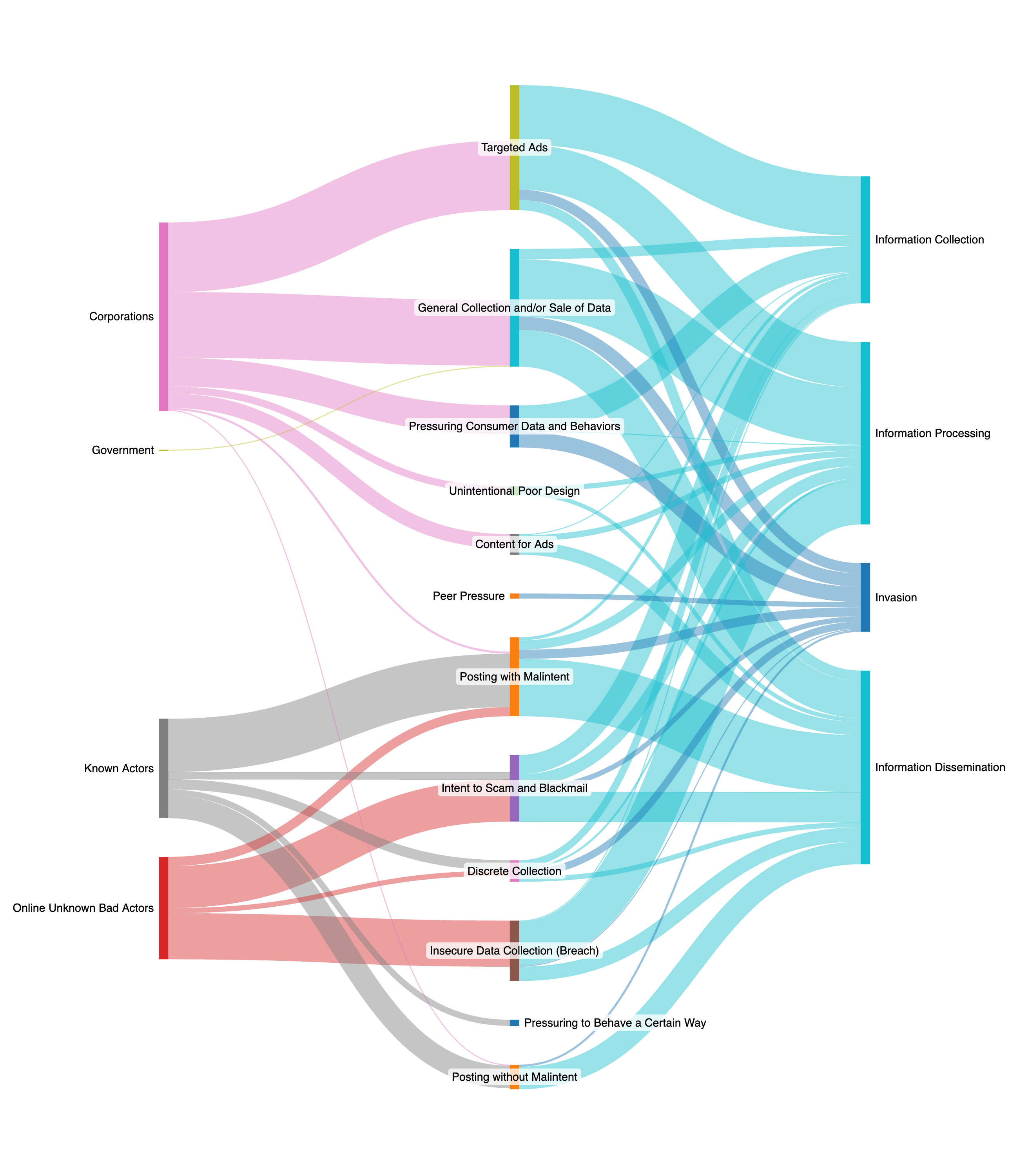}
    \caption{Mapping of involved actors to underlying motives to the incident type that occurred.}
    \label{fig:app1}
\end{figure}

\begin{figure}[htp!]
    \centering
    \includegraphics[width=0.75\linewidth]{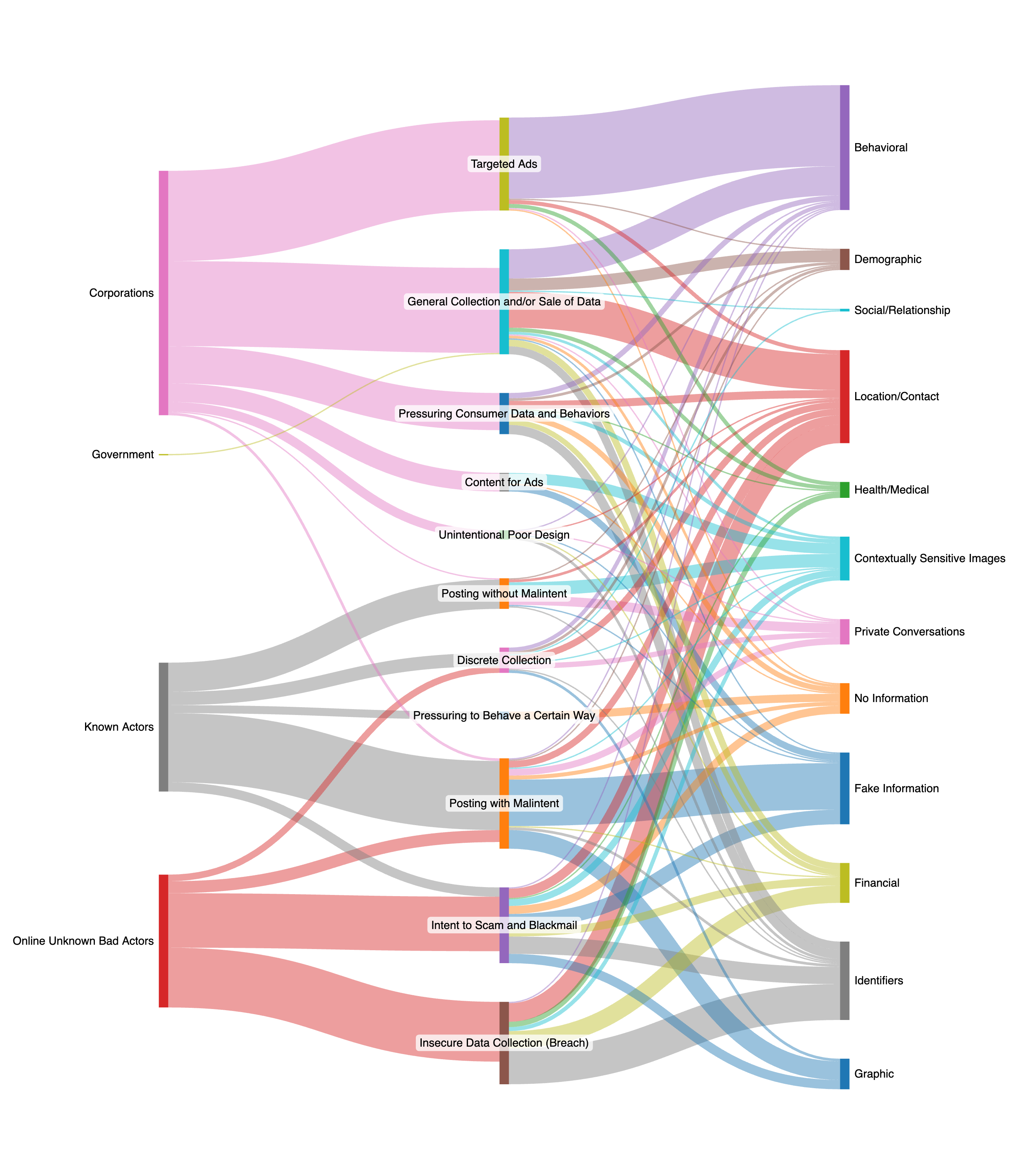}
    \caption{Mapping of involved actors to underlying motives to the information type that was involved.}
    \label{fig:app2}
\end{figure}

\end{document}